# An etched multilayer as a dispersive element in a curved-crystal spectrometer: implementation and performance

P. Jonnard, K. Le Guen, J.-M. André
Laboratoire Chimie Physique – Matière et Rayonnement, UPMC Univ Paris 06, CNRS UMR 7614, 11 rue Pierre et Marie Curie, F-75231 Paris cedex 05, France

J.-R. Coudevylle, N. Isac
Institut d'Electronique Fondamentale, CNRS UMR 8622, Bât. 220, Université Paris Sud XI, F-91405 Orsay Cedex, France

**Abstract**
Etched multilayers obtained by forming a laminar grating pattern within interferential multilayer mirrors are used in the soft x-ray range to improve the spectral resolution of wavelength dispersive spectrometers equipped with periodic multilayers. We describe the fabrication process of such an etched multilayer dispersive element, its characterizationthrough reflectivity measurement and simulations, and its implementation in a high-resolution Johann-type spectrometer. The specially designed patterning of a Mo/$B_4$C multilayer is found fruitful in the range of the C K emission as the diffraction pattern narrows by a factor 4 with respect to the non-etched structure.This dispersive element with an improved spectral resolution was successfully implemented for electronic structure study with an improved spectral resolution by x-ray emission spectroscopy. As first results we present the distinction between the chemical states of carbon atoms in various compounds, such as graphite, SiC and $B_4$C, bythe different shape of their C K emission band.

**Keywords:**dispersive element; multilayer; soft x-ray spectrometer; x-ray emission spectroscopy

**Short title:**An etched multilayer setup in a Johann-type spectrometer

# Introduction

The spectral resolution of a wavelength dispersive spectrometer is generally limited by the width of the diffraction pattern of the dispersive element. To study the soft x-ray range, *i.e.* for photon energies between 50 and 1000 eV, the spectrometers can be equipped with periodic multilayers[1]–[3], called multilayer mirrors (MM) in the following. However, due tostrong absorption occurring in this range, onlya small number of periods contributes to the diffraction process and the diffraction pattern of the multilayer is wide. Typically, in the range of the C K emission around 280 eV, the diffraction pattern can be as large as 25–30 eV. Under these conditions, the spectral resolution is poor and spectrometers can only be used to measure the intensity of characteristic lines. However, a high spectral resolution spectrometer is needed to minimize the interferences between close lines, to improve the detection limit and to observe the real shape of emission bands in order to determine the chemical state of the emitting elements. The use of narrow band MM[4], in fact working at a high diffraction order, could also be imagined at the detriment of obtaining a poor reflectance.

Since the pioneer work of André *et al.*[5]–[8], it has been shown theoretically and experimentally that it is possible to improve the spectral resolution by etching the multilayer according to the pattern of a laminar grating[8]–[15], thus obtaining an etched multilayer (EM). We present in Figure 1 the scheme of such a structure characterized by two periods: the period of the multilayer in which absorbing and spacer layers are alternating, and the period of the gratingwhere grooves and multilayer bars succeed each other.The ratio of the grating period to the width of the multilayer bar is defined as the $\Gamma$ parameter. Because of the voids generated by the grooves in the etched structure, the radiation penetrates deeper within the structure, enabling an efficient diffraction process, which leads to a significant decrease of the width of the diffraction pattern with respect to the non-etched structure. It has been shown that this reduction can be of the order of $1/\Gamma$ [15][16] and especially can be achieved with a minimal reflectivity loss provided the period of the grating is no larger than 1 μm[11].

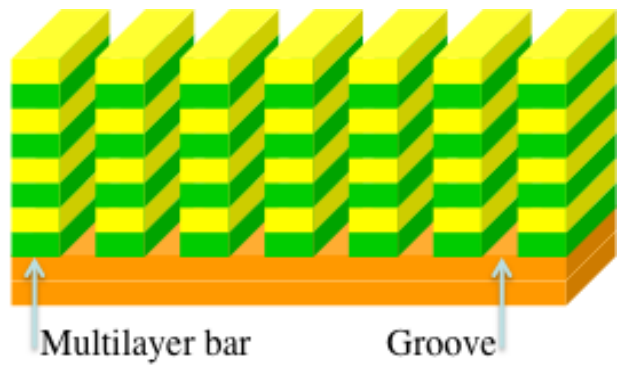

Multilayer bar     Groove

Figure 1: Scheme of a multilayer grating. The multilayer period is the sum of the thicknesses of the absorbing (yellow) and of the spacer (green) layers. The grating period is the sum of the widths of the multilayer bar and of the groove.

The etching process is done in clean rooms following the recipes of microelectronics through photolithography[17] or nano-imprint[15]. Electron lithography is avoided because it is time consuming and therefore not suited for patterning large area samples. EM having a period as small as 200 nm and a 60 nm-wide multilayer bar, thus a Γ ratio of 0.3, has already been produced[15]. Another way to obtain similar structures is to deposit a multilayer onto a grating substrate[13].

One difficulty to fabricate such structures is the need for a high aspect ratio (the ratio of the height to the width). Indeed, the multilayer height is generally a few micrometers (typically 200 periods of 10 nm thickness), while following etching the width of the multilayer bar is only some hundreds or even tens of nanometers. These structures are therefore fragile and cannot withstand large mechanical stress. This is the main reason why, to our knowledge, EM have never been implemented in curved-crystal spectrometers, for fear that the curvature of the structure will induce the delamination of the multilayer bars from the substrate. In this paper we show that it is possible to implement an EM in a Johan-type curved-crystal spectrometer and we use it to identify the chemical state of an element emitting in the soft x-ray range.

## Preparation and characterization of the multilayer grating

The EM was prepared from a Mo/$B_4C$ MM deposited using magnetron sputtering. The structure of the multilayer is the following: Si substrate / 5 nm Cr / [2.0 nm Mo / 3.9 nm $B_4C$]$_{X200}$. This MM is designed to present a high reflectivity around 180 eV or 6.7 nm (B K emission range). The chromium layer on the substrate has a double role: to obtain a good adhesion between the substrate and the multilayer and to stop the etching process just at the substrate surface. The process of the MM is divided into the following steps:

- deposition of a $SiO_2$ layer;
- deposition of a UVIII 0.9 resist by spin coating;
- insolation at 248 nm through a mask having the desired pattern;
- development of the resist;
- deposition of an Al film to protect the sample from subsequent etching;
- lift-off of the UVIII 0.9 / Al zones;
- fluorine reactive ion etching;

- removal of the SiO$_2$ / Al zones in a HF bath.

More details on the process can be found in[18]. The patterned zone has a size of 10 mm x 10 mm. A scanning electron image of the EMprior to the last step is shown in Figure 2. Thus the multilayer bars are still covered by a silica film that induces a charge effect and prevents obtaining a high quality image. The period of the grating is 1 µm and the width of the multilayer bar 180 nm, leading to a Γ ratio equal to 0.18. It is observed that the bottom of the grooves is clean and that the width of the multilayer bar is not constant over the height due to over-etching. However, this inhomogeneity should not hamper the performances of the EM in terms of bandpass narrowing, as the main role of the grooves is to create some voids within the MM stack.Nevertheless a decrease of reflectivity could be expected to this non-regular in-depth profile. This question remains open and is under study by simulation.

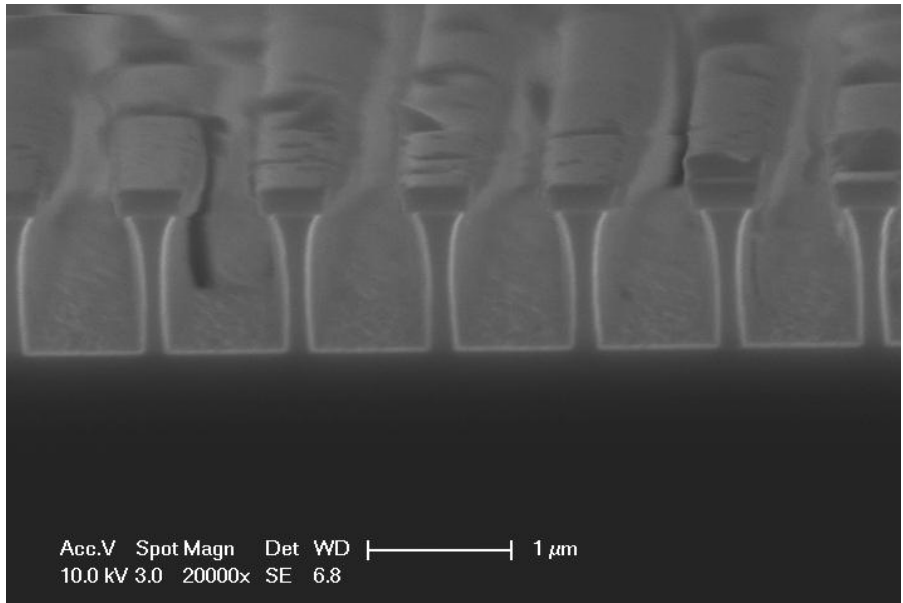

Figure 2: Scanning electron microscopy image of the etched multilayer.

The prepared EM has been characterized in the B K and C K (around 280 eV, 4.5 nm) ranges on the BEARbeamline[19]at the Elettra synchrotronradiation facility. For both ranges, the reflectivity was measured as a function of the grazing angleand the photon energy, as shown in Figure 3.The curves are obtained at the first Bragg order of the MM and at the zeroth diffraction order of the grating. Some oscillations of the reflectance are observed for the EM and not for the MM which present otherwise rather smooth profiles. This is due to coupling of the diffraction orders (±1, ±2, …)of the gratingas accounted for by the modal theory[7]. The values of the width of the reflectivity curves and of the reflectancesof both MM and EMare collectedin Table 1 as well as the relative variations of these values from MM to EM.

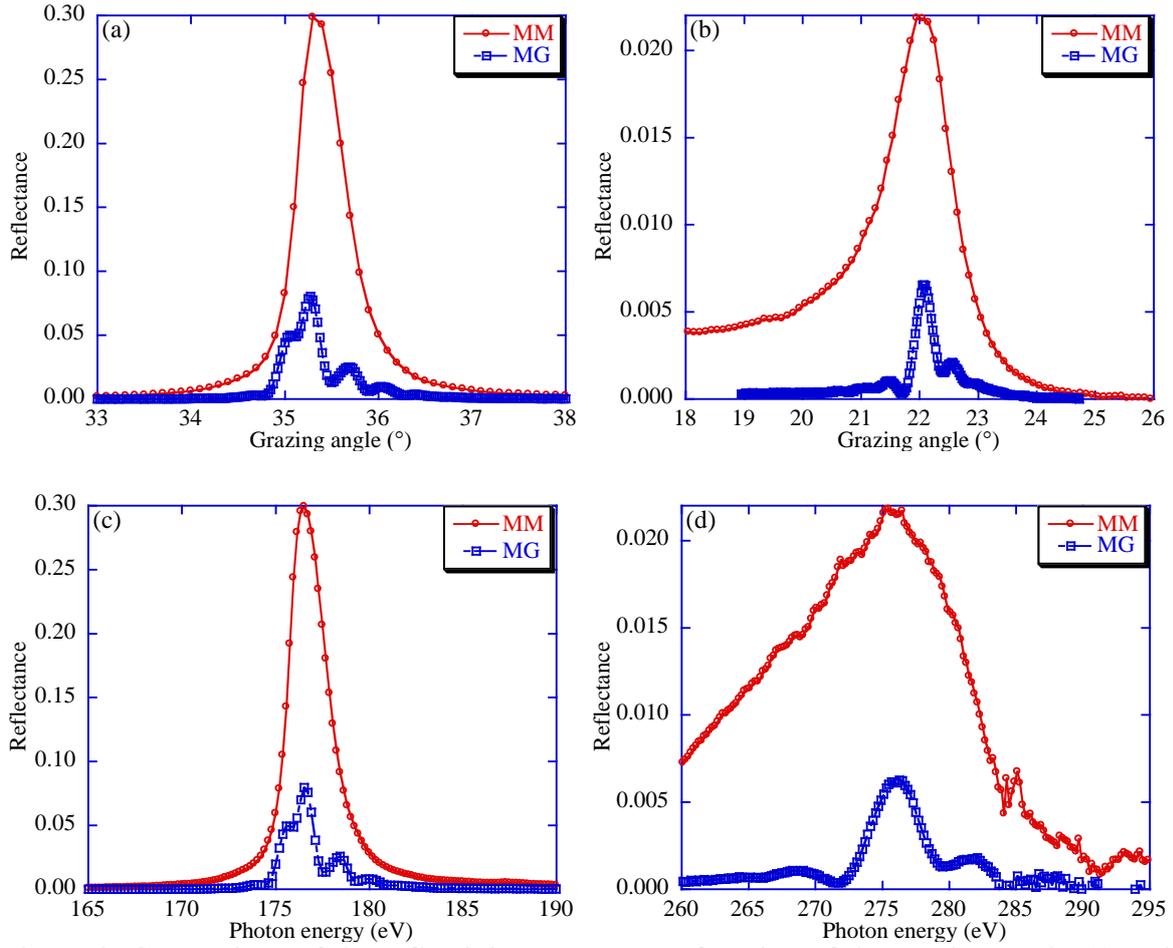

Figure 3: Comparison of the reflectivity curves, as a function of the grazing angle (a) and (b) and as function of the photon energy (c) and (d) measured for the Mo/B$_4$C multilayer mirror (MM, red curves) and the etched multilayer (EM, blue curves): (a) in the B K range at a photon energy of 180 eV, (b) in the C K range at a photon energy of 277 eV, (c) in the B K range at a grazing angle of 35.10° and (d) in the C K range at a grazing angle of 22.05°.

| Range | B K | | C K | |
|---|---|---|---|---|
| *Grazing angle* | R (%) | L$_{1/2}$ (°) | R (%) | L$_{1/2}$ (°) |
| MM | 29.9 | 0.59 | 2.18 | 1.40 |
| EM | 8.0 | 0.41 | 0.65 | 0.34 |
| MM/EM | 3.7 | 1.4 | 3.4 | 4.1 |
| *Photon energy* | R (%) | L$_{1/2}$ (eV) | R (%) | L$_{1/2}$ (eV) |
| MM | 29.9 | 2.33 | 2.18 | 14.1 |
| EM | 8.0 | 1.78 | 0.63 | 4.2 |
| MM/EM | 3.7 | 1.3 | 3.5 | 3.5 |

Table 1: Comparison as a function of the grazing angle and of the photon energy, of the reflectance (R) and full width at half-maximum (L$_{1/2}$) of the reflectivity curves, measured for the multilayer mirror (MM) and the etched multilayer (EM) in the B K and C K ranges.

From the examination of data in Table 1 it appears that results obtained as a function of the grazing angle or of the photon energy are almost equivalent. In the C K range, the width decreases by a factor of about 4 when going from the MM to the EM, demonstrating the successful patterning of the sample. However, in the B K range for which the MM was designed, the width decrease is only 30 to 40%. At present time, this behaviour difference of the width evolution as a function of the range is not fully understood. However, in the B K range a shoulder is observed close to the main peak, at -0.2° on the curve as a function of the angle. As stated above, this feature is due to modal interferences originating from the diffraction by the lamellar grating[7]. Since its intensity is higher than half the peak intensity, it is included in the measurement of the full width at half-maximum. If we consider only the half-width at half-maximum towards the high Bragg angles in order to avoid this grating structure, then the narrowing with respect to the MM would be a factor 3, *i.e.* much closer to the value obtained in the C K range.The occurrence of the modal interference structure is problematic and should be avoided.Working with anEM whose grating period is smaller than the present one, would minimize this annoying interference between grating and multilayer diffraction peaks by allowing to work in the single-order regime[16].

Regarding the reflectance, it decreases by a factor slightly less than 4 when going from the MM to the EM whatever the range, see Table 1. Two reasons for this large variation have been identified:

- the $\Gamma$ ratio of the grating is only 0.18 due to an over-etching; this is not the optimum value which is in the 0.25 – 0.35 range; a $\Gamma$ ratio in this range would have limited the reflectivity decrease to only 30%[11];
- the first step of the patterning process is the deposition of a $SiO_2$ layer requiring that the sample stays at 300°C during at least 15 minutes; this condition leads to a degradation of the MMstructural quality, as we have checked from reflectivity measurements in the Cu L (930 eV, 1.33 nm) range with the MONOX spectro-goniometer[20]; preliminary silica depositions at 150°C show that the MM reflectance is unaffected by this temperature annealing.

Let us recall that the non-regular profile of the multilayer bars should also affect the reflectance as previously mentioned.

We show in Figure 4 the comparison between the experiment and two simulations of the reflectivity curves obtained as a function of the angle and in the B K and C K ranges. In the first simulation, we use the framework of the modal theory (MT) modified for periodic multilayer structures[7][21]. For this simulation, 17 diffraction orders of the grating are introduced in the calculation. In the second simulation, the coupled-wave approach (CWA) in the single-order regime[16] is used.The interest of the MT model is that it is rigorous and the results are only affected by the truncation in the number of modes retained in the computation, whereas the interest of the CWA model resides in its short computing time. In both cases, perfect multilayer and grating are considered, *i.e.* no interfacial roughness and no interdiffusion in the multilayer and no lateral or vertical roughness in the grating whose grooves and bar are perfectly rectangular.

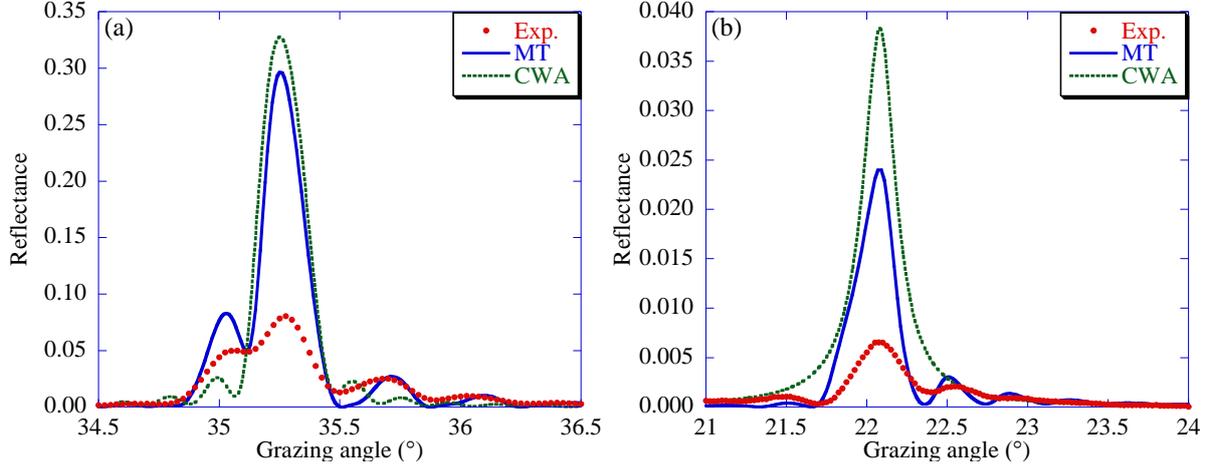

Figure 4: Comparison of the reflectivity curves obtained as a function of the angle and in the (a) B K and (b) C K ranges between the experiment (dots), simulation with the modal theory model (solid line) and simulation with coupled-wave approach in the single-order regime (dotted line).

It is observed that both simulations give an efficiencylarger than the experimental one. Some of the possible reasons have been discussed above. The MT and the CWA in the single-mode regime do not give the same peak reflectivity in our present case: the CWA in this regime is likely not valid in our conditions but this question needs to be clarify and the validity domain of the CWA has to be determined.

Both models give almost the same bandwidth demonstrating again the resolution improvement introduced by the EM, by a factor about 3 with respect to the experimental width of the MM in the B K range and by a factor 5 in the C K range. The positions of the experimentally observed oscillations at each side of the main diffraction peak are well reproduced in the MT model but not in the CWA model. This is because only the zeroth order is taken into account by the CWA modelin the single-order regime. A discrepancy occurs between experiment and simulation with the MT model regarding the intensities of the oscillations relative to one of the main peak. This is not well understood yet, but may be due to the non-rectangular shape of the multilayer bars as seen in the SEM image in Figure 2.In the MT model used in the B K range, the oscillation at -0.2° from the main peak is less intense than half the intensity of the peak and thus the bandwidth narrowing by a factor 3 with respect to the experiment can be observed.

## Setup in a curved-crystal spectrometer

The fabricatedEM has been implemented in the crystal holder of the IRIS[22]high-resolution x-ray spectrometer. This is a high-resolution Johann-type bent-crystal spectrometer in which we previously replaced crystals by periodic multilayers to study the emissions occurring in the soft x-ray range[3]. The high resolution is obtained through the use of a small opening of the dispersive element (crystal, multilayer or etched multilayer), 10 mm x 8 mm, combined to a large radius of curvature of the crystal, 500 mm. Indeed, part the instrumental broadening scales with the square of the opening size[22] and the large radius allows a wide dispersion of the radiation.The other advantage of the large radius is that itensures a limited stress within the stack and thus minimizes the risk of delamination of the multilayer from its substrate and therefore we have been able to work successfully with the EM.

We have analyzedby x-ray emission spectroscopy the C K emission of three different samples: a bulk graphite target andSiC and $B_4C$ thin films. In the IRIS apparatus, samples are placed outside the Rowland circle, at about 600 mm from it. This setup allows collecting the radiation emitted by large samples.The excitation was done by using an 8 keV electron beam whose spot on the sample is about 1 $cm^2$. A Geiger-type detector was used for the x-ray detection.Let us note that, despite the low reflectance of the EM, 0.6%, reliable spectrawere obtained owing to the high luminosity of the spectrometer and the high carbon content within the studied compounds.

The comparison of the emission bands of the three samples is given in Figure 5.The bandwidth is about 10 eV wide, whereas if we had used the MM this width would have been slightly less than 20 eV[11][14].This is an improvement of the resolution by a factor of only 2. Indeed, even if we expect a decrease by a factor 4 of the bandwidth of the dispersive element, this factor is not the only one to take into account to estimate the width of an observed emission. In fact, the emission bands of the light elements have relatively large natural widths, about 10 eV.When combined quadratically with the experimental broadening, mainly due to the diffraction pattern of the EM, it gives the factor 2 between the emission widths observed with the MM and the EM.For narrower lines, such as atomic lines or the ones observed in plasmas, the improvement factor of the resolution would have been much closer to 4.

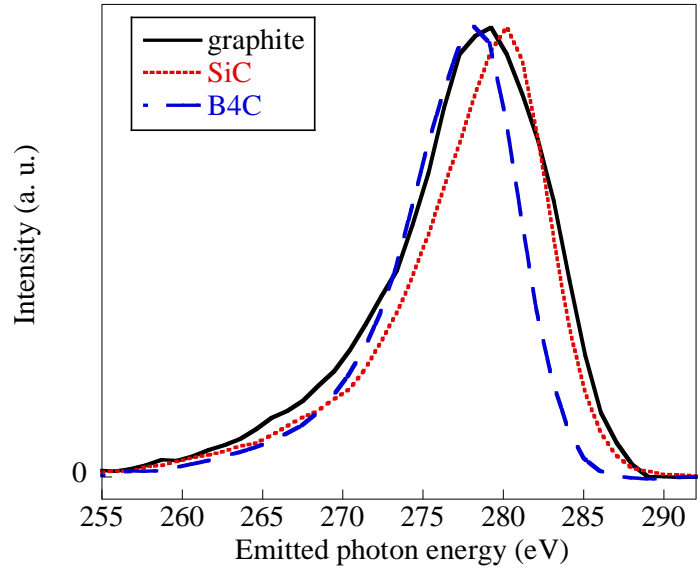

Figure 5: Comparison of the C K emission bands of graphite (solid line), silicon carbide (dotted line) and boron carbide (dashed line) obtained with the EM set up as the dispersive element in a curved-crystal spectrometer.

The improved resolution given by the EM enables us to discriminate between the different chemical states of the C emitting atoms. Indeed, each emission band has its own position of the maximum, width and shape. Thus it is worth using this kind of EM for electronic structure study and not only for intensity measurements for the purpose of quantitative analysis. This opportunity had already been demonstrated in our previous EM characterization study[12], by distinguishing the chemical state of boron carbide and cellulose.

In addition, we have compared the C K emissionofgraphite and $B_4C$ obtained with the EM implemented in IRIS[22]and with a grating spectrometer[23].For this last kind of spectrometer, absorption does not limit the spectral resolution,as the diffracting structures are at the surface of the grating. Then, much better resolution can be reached in the soft x-ray and extreme UV ranges. For example,a resolution of a few thousands (E/ΔE) can be obtained in our photon energy range[24]. The comparison presented in Figure 6shows that, with respect to the grating spectrometer, the EM still provides a larger bandwidth. However, despite the lower resolution, particularly for graphite,the features of the valence band can still be observed with the EM. With the MM, the C K emission band would have been observed as a broad and rather symmetrical peak[12][14].

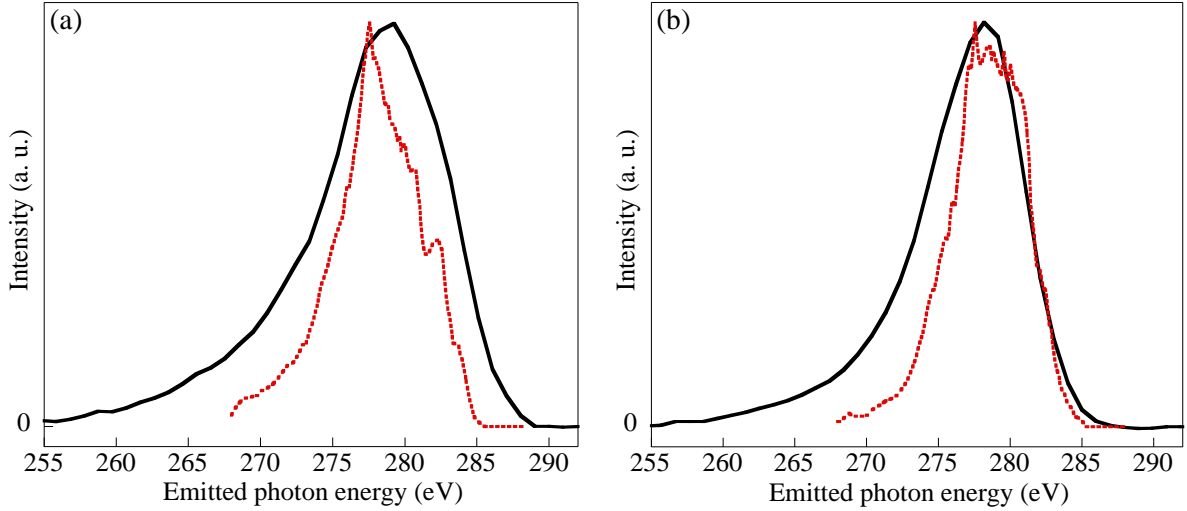

Figure 6: Comparison of the C K emission bands of graphite (a) and boron carbide (b) obtained with the EM (solid line) setup in the curved-crystal spectrometer and with a grating spectrometer[23] (dotted line).

## Conclusion

We have characterized and simulated a new fabricatedmultilayer grating and shown that it improves the spectral resolution by a factor 4 in the C K range, with respect to the non-etched multilayer. The reflectance of such a structure is not yet satisfactory but will be improved by a better control of the patterning process of the MM.Another mean to improve the reflectance in the C K range would be to etch multilayers specifically designed for this range, such as W/C or MT/C (MT = Fe, Co or Ni) multilayers. However, this could be envisaged only after a serious examination of the possibility to apply the developed process to such MM.

The produced EM was implemented as the dispersive element of a curved-crystal spectrometer and operated successfully to distinguish the C K emission bands originating fromdifferent compounds. Thus, this kind of structure can be used to study the electronic structure of various compounds. Finally, EMappearsto be a good compromise between MM set up in bent-crystal spectrometers and grating spectrometers, giving a spectral resolution intermediate between those of the MM and the grating spectrometer, while using the same kind of spectrometer as the MM. This is important for people who are operating only crystal spectrometers as in electron microprobes and cannot afford to also design, build and use a grating spectrometer.

## Acknowledgments

Dr. R. Benbalagh from LCPMR is acknowledged for helpful discussions. The team of the BEAR beamline from ELETTRA is thanked for its help during the reflectivity measurements.X.-F. Zhu is thanked for her experimental help during the x-ray emission measurements.Incoatec is thanked for providing us with the Mo/$B_4$C multilayers.

## References

[1] J. Underwood, T. Barbee Jr, others, *Appl. Opt.* **1981**, *20*, 3027–3034.
[2] T.W. Barbee, *Phys. Scr.* **1990**, *T31*, 147–153.
[3] C. Hombourger, P. Jonnard, J.-M. André, J.-P. Chauvineau, *X-Ray Spectrom.* **1999**, *28*, 163–167.
[4] T. Feigl, S. Yulin, N. Benoit, N. Kaiser, *Microelectronic Engineering* **2006**, *83*, 703–706.
[5] H. Berrouane, J.-M. André, R. Barchewitz, C.K. Malek, R. Rivoira, *Opt. Commun.* **1990**, *76*, 111–115.
[6] B. Pardo, J.-M. André, A. Sammar, *J. Opt.* **1991**, *22*, 141–148.
[7] A. Sammar, J.-M. André, B. Pardo, *Opt. Commun.* **1991**, *86*, 245–254.
[8] A.-E. Sammar, J.-M. André, M. Ouahabi, B. Pardo, R. Barchewitz, *C. R. Acad. Sci., Ser. 2,* **1993**, *316*, 1055–1060.
[9] K. Krastev, F. Leguern, K. Coat, R. Barchewitz, J.-M. André, M.-F. Ravet, E. Cambril, F. Rousseaux, P. Davi, *Ann. Phys.* **1997**, *22*, 149–150.
[10] V.V. Martynov, Y. Platonov, *Adv. X-Ray Anal.* **2002**, *45*, 402–408.
[11] R. Benbalagh, J.-M. André, R. Barchewitz, P. Jonnard, G. Julié, L. Mollard, G. Rolland, C. Rémond, P. Troussel, R. Marmoret, E.O. Filatova, *Nucl. Instrum. Methods Phys. Res. Sect. A* **2005**, *541*, 590–597.
[12] P. Jonnard, R. Benbalagh, J.-M. André, *Microsc. Microanal.* **2007**, *13*, 162–163.
[13] M. Störmer, J.-M. André, C. Michaelsen, R. Benbalagh, P. Jonnard, *J. Phys. D: Appl. Phys.* **2007**, *40*, 4253–4258.
[14] P. Jonnard, K. Le Guen, J.-M. André, *X-Ray Spectrom.* **2009**, *38*, 117–120.
[15] R. van der Meer, B. Krishnan, I.V. Kozhevnikov, M.J. De Boer, B. Vratzov, H.M.J. Bastiaens, J. Huskens, W.G. van der Wiel, P.E. Hegeman, G.C.S. Brons, K.-J. Boller, F. Bijkerk, *Proc. SPIE* **2011**, *8139*, 81390Q.
[16] I.V. Kozhevnikov, R. van der Meer, H.M.J. Bastiaens, K.-J. Boller, F. Bijkerk, *Opt. Express* **2010**, *18*, 16234–16242.
[17] R.P. Seisyan, *Tech. Phys.* **2011**, *56*, 1061–1073.
[18] R. Benbalagh, *Monochromateurs multicouches à bande passante étroite et à faible fond continu pour le rayonnement X–UV*, PhD Thesis, Université Pierre et Marie Curie, **2003**.
[19] S. Nannarone, F. Borgatti, A. DeLuisa, B.P. Doyle, G.C. Gazzadi, A. Giglia, P. Finetti, N. Mahne, L. Pasquali, M. Pedio, G. Selvaggi, G. Naletto, M.G. Pelizzo, G. Tondello, *AIP Conf. Proc.* **2004**, *705*, 450–453.
[20] J.-M. André, A. Avila, R. Barchewitz, R. Benbalagh, R. Delaunay, D. Druart, P. Jonnard, H. Ringuenet, *Eur. Phys. J. Appl. Phys.* **2005**, *31*, 147–152.
[21] K. Krastev, J.-M. André, R. Barchewitz, *J. Opt. Soc. Am. A* **1996**, *13*, 2027–2033.
[22] C. Bonnelle, F. Vergand, P. Jonnard, J.-M. André, P.-F. Staub, P. Avila, P. Chargelègue, M.-F. Fontaine, D. Laporte, P. Paquier, A. Ringuenet, B. Rodriguez, *Rev. Sci. Instrum.* **1994**, *65*, 3466–3471.
[23] J. Iihara, Y. Muramatsu, T. Takebe, A. Sawamura, A. Namba, T. Imai, J.D.

Denlinger, R.C.C. Perera, *Jpn. J. Appl. Phys.* **2005**, *44*, 6612–6617.

[24] V.N. Strocov, T. Schmitt, U. Flechsig, L. Patthey, G.S. Chiuzbăian, *J. Synchr. Rad.* **2011**, *18*, 134–142.